\documentclass[12pt]{article}
\newlength{\mytopmargin}
\newlength{\myleftmargin}
\usepackage{amsmath,amsthm,amssymb,amsbsy,epsfig,graphicx,color,multicol,subfigure}
\usepackage{array,calc}
\usepackage[enableskew]{youngtab}
\usepackage{rotating}
\usepackage{young,epic}
\usepackage{a4wide,bm}
\usepackage{url}
\usepackage{hyperref}

\newtheorem{prop}{Proposition}

\usepackage{amsmath,amsfonts,amssymb}
\usepackage{epsf}
\usepackage{eepic}

\usepackage{graphicx}

\newcommand\psymmU{%
\begin{picture}(1,1)(0,0)%
\allinethickness{0.5pt}%
\path(0,0)(0,1)(1,1)(1,0)(0,0)%
\end{picture}}
\newcommand\psymmUU{%
\begin{picture}(1,1)(0,0)%
\allinethickness{0.5pt}%
\path(0,0)(0,1)(1,1)(1,0)(0,0)%
\put(0.5,0.5){\makebox(0,0){$\cdot$}}%
\end{picture}}
\newcommand\psymmO{%
\begin{picture}(1,1)(0,0)%
\allinethickness{0.5pt}%
\path(0,0)(0,1)(1,1)(1,0)(0,0)%
\path(0,0)(1,1)%
\end{picture}}
\newcommand\psymmS{%
\begin{picture}(1,1)(0,0)%
\allinethickness{0.5pt}%
\path(0,0)(0,1)(1,1)(1,0)(0,0)%
\path(1,0)(0,1)%
\end{picture}}
\newcommand\psymmu{%
\begin{picture}(1,1)(0,0)%
\allinethickness{0.5pt}%
\path(0,0)(0,1)(1,1)(1,0)(0,0)%
\path(0,0)(1,1)%
\path(0,1)(1,0)%
\end{picture}}

\newbox\tsymmUbox
\newbox\tsymmUUbox
\newbox\tsymmObox
\newbox\tsymmSbox
\newbox\tsymmubox
\setbox\tsymmUbox =\hbox{\kern0.75pt\setlength{\unitlength}{6pt}\psymmU \kern0.75pt}

\setbox\tsymmUUbox=\hbox{\kern0.75pt\setlength{\unitlength}{6pt}\psymmUU\kern0.75pt}
\setbox\tsymmObox =\hbox{\kern0.75pt\setlength{\unitlength}{6pt}\psymmO \kern0.75pt}
\setbox\tsymmSbox =\hbox{\kern0.75pt\setlength{\unitlength}{6pt}\psymmS \kern0.75pt}
\setbox\tsymmubox =\hbox{\kern0.75pt\setlength{\unitlength}{6pt}\psymmu \kern0.75pt}

\newbox\symmUbox
\newbox\symmUUbox
\newbox\symmObox
\newbox\symmSbox
\newbox\symmubox
\setbox\symmUbox =\hbox{\kern0.75pt\setlength{\unitlength}{4.5pt}\psymmU \kern0.75pt}
\setbox\symmUUbox=\hbox{\kern0.75pt\setlength{\unitlength}{4.5pt}\psymmUU\kern0.75pt}
\setbox\symmObox =\hbox{\kern0.75pt\setlength{\unitlength}{4.5pt}\psymmO \kern0.75pt}
\setbox\symmSbox =\hbox{\kern0.75pt\setlength{\unitlength}{4.5pt}\psymmS \kern0.75pt}
\setbox\symmubox =\hbox{\kern0.75pt\setlength{\unitlength}{4.5pt}\psymmu \kern0.75pt}

\def\symmO{{\copy\symmObox}}
\def\symmS{{\copy\symmSbox}}

\begin{document}
%

\title{Painleve II in random matrix theory and related fields}
\author{Peter J. Forrester and Nicholas S. Witte}
\date{}
\maketitle
\noindent
\thanks{\small Department of Mathematics and Statistics, 
The University of Melbourne,
Victoria 3010, Australia email:  p.forrester@ms.unimelb.edu.au \: \: nsw@ms.unimelb.edu.au 
}

\begin{abstract}
\noindent We review some occurrences of Painlev\'e II transcendents in the study of two-dimensional Yang-Mills theory,
fluctuation formulas for growth models, and as distribution functions within random matrix theory. We first discuss
settings in which the parameter $\alpha$ in the Painlev\'e equation is zero, and the boundary condition is that of the
Hasting-MacLeod solution. As well as expressions involving the Painlev\'e transcendent itself, one encounters the
sigma form of the Painlev\'e II equation, and Lax pair equations in which the Painlev\'e transcendent occurs as
coefficients. We then consider settings which
give rise to general $\alpha$ Painlev\'e II transcendents. In a particular random matrix setting, new results 
for the corresponding boundary conditions
in the cases $\alpha = \pm 1/2$, 1 and 2 are presented.
\end{abstract}

\section{Introduction}
A Painlev\'e II transcendent refers to a solution of the Painlev\'e II nonlinear differential equation
\begin{equation}\label{1.1}
q_\alpha'' = t q_\alpha + 2 q_\alpha^3 - \alpha.
\end{equation}
For special values of the parameter $\alpha$, in particular $\alpha = 0$,
and special boundary conditions determining a particular solution of (\ref{1.1}),
Painlev\'e II transcendents are prevalent in random matrix theory. They come about in the so-called double scaling limit,
taken in the study of matrix models (see e.g.~\cite{Bl11}), as the scaling limit of certain probabilities specified as averages
over unitary, orthogonal and unitary symplectic ensembles, as well as the scaling limit of the gap probabilities for ensembles
of Hermitian matrices at the soft edge, when the leading order eigenvalue density goes to zero like a square root (these latter
two occurrences will be reviewed below). As also to be reviewed below, it is furthermore the case that certain Lax pairs or Painlev\'e II transcendents arise
in the context of the study of the scaling limit of some generalizations of the above settings which involve an additional parameter.

Our primary interest in this paper is to highlight the occurrence in random matrix theory of (\ref{1.1}) for general $\alpha$. However, we take
the opportunity to also review more generally specific settings in random matrix theory and its applications leading to Painlev\'e II transcendents.
In this regard, low-dimensional field theories, which give rise to random matrix (or random matrix-like) averages, will be considered first. These
very same averages appear in the study of some probabilities associated with certain combinatorial models and growth processes, so these
will be considered next in Section 2.3. In Section 2.4 we turn to the problem of soft edge gap probabilities in Hermitian random matrix ensembles
with unitary, orthogonal and symplectic symmetry. The corresponding formulas in terms of a Painlev\'e II transcendent for the probability of no
eigenvalue in $(s,\infty)$ are precisely the same as those appearing as limiting probabilities of the combinatorial models from Sections 2.2 and 2.3. An
explanation for this involving first a transformation identity between probabilities, then an analysis of the hard-to-soft edge transition in random
matrix theory as given in \cite{BF03} is revised. In Section 2.5 we describe settings, within combinatorial and growth models, and within random
matrix theory at the soft edge, which give rise to Lax pairs for a Painlev\'e II transcendent.

The  Painlev\'e II transcendents in the above problems either literally satisfy the case $\alpha = 0$ of (\ref{1.1}), or can be mapped to that case.
In Section 3 we turn our attention to a setting in random matrix theory at the soft edge in which $\alpha$ is a continuous variable.
After presenting some review material we proceed to give new results in the cases $\alpha = \pm 1/2$, 1 and 2.

\section{A brief survey of Painlev\'e II with $\alpha = 0$ in random matrix theory and applications}
\setcounter{equation}{0}
\subsection{Two-dimensional Yang-Mills gauge theory}
In two-dimensions QCD without quarks is a pure Yang-Mills gauge theory. The setting involves a two-dimensional orientable manifold $\mathcal M$ of
volume form $\sqrt{g}$, containing at each point $\vec{x}$ a gauge field $(A_1(\vec{x}), A_2(\vec{x}))$. The latter are taken to be $N \times N$ matrices
from the Lie algebra of an appropriate gauge group $G$, and from these field strengths are defined by
\begin{equation}\label{FA}
F_{\mu \nu} = \partial_\mu A_\nu -  \partial_\nu A_\mu  + i [A_\mu, A_\nu], \qquad (\mu,\nu = 1,2).
\end{equation}
A fundamental property of the field strengths is that under the local gauge transformation
$$
A_\mu \mapsto S^{-1}(\vec{x}) A_\mu S(\vec{x}) - i  S^{-1}(\vec{x})  \partial_\mu S(\vec{x}), \quad  S(\vec{x}) \in G,
$$
they transform according to $F_{\mu \nu} \mapsto   S^{-1}(\vec{x})  F_{\mu \nu} S(\vec{x})$.
This in turn demonstrates the gauge invariance of the Yang-Mills action
$$
I_{\rm YM} = {1 \over 4 \lambda^2} \int_{\mathcal M} {\rm Tr} (F^{\mu \nu} F_{\mu \nu}) \sqrt{g} \,
d^2 \vec{x},
$$
where $1/\lambda^2$ is a coupling constant, and also of the partition function
\begin{equation}\label{ZG}
Z_{\mathcal M}(\lambda;G) = \int [\mathcal D A_\mu] e^{- I_{\rm YM}}.
\end{equation}
The notation $G$ in (\ref{ZG}) denotes the particular gauge group under consideration, while $ [\mathcal D A_\mu] $ denotes a suitable measure on
the fields.

One approach to the computation of (\ref{ZG}) is to discretize the manifold and also the action \cite{Mi75}. In two-dimensions the manifold can be triangulated
into basic units called plaquettes. On each edge $L$ of the triangulation, and with gauge group $G$, one associates an element $G_L \in G$ and defines
the corresponding lattice regularized partition function according to
\begin{equation}\label{23}
Z_{\mathcal M}^{\rm lattice} = \int \prod_L (d G_L) \prod_{\rm plaquettes} Z_P(G_P),
\end{equation}
where $(d G)$ denotes the Haar measure on $G$.
Here $G_P =  \prod_{L \in \rm plaquettes} G_L$ is the Wilson loop product \cite{Wi74}. Taking $Z_P$ as the Wilson action,
\begin{equation}\label{24'}
 Z_P(G_P) = \exp \Big ( b N {\rm Tr} (G_P + G_P^\dagger) \Big ),
 \end{equation}
 which in the case that $G$ is a classical group is a valid choice since it reduces to $e^{- I_{\rm YM}}$ in the continuum limit, one sees that
 (\ref{23}) factorizes as
 \begin{equation}\label{25}
 Z_{\mathcal M}^{\rm lattice} = \Big ( \Big \langle e^{b N {\rm Tr} (G_L + G_L^\dagger)} \Big \rangle_{G_L \in G} \Big )^M,
 \end{equation}
 where $M$ is the number of plaquettes. Hence we have the result that $ (Z_{\mathcal M}^{\rm lattice})^{1/M}$ is given by an explicit matrix
 integral. We will revise below how a certain double scaling limit --- taking $N \to \infty$ but also tuning $b$ as a function of $N$ ---
 leads to a Painlev\'e II transcendent. In the context of the present problem the double scaling limit is motivated by the desire to
 relate two-dimensional lattice QCD to string theory; see e.g.~\cite{PS10}.
 
 The choice of action (\ref{24'}) is not unique if the only requirement is that it reduces to  $e^{- I_{\rm YM}}$ in the continuum limit.
 In addition to requiring that $Z_P(G_P)$ reduces to the Yang-Mills action, following \cite{Mi75} one can also require that fusing two adjacent
 plaquettes $P_1$ and $P_2$, and integrating out the group element associated with the common edge give back the action
 $Z_{P_1+P_2}$,
 \begin{equation}\label{ZZ}
 \int (d G_3) \, Z_{P_1}(G_1 G_2 G_3) Z_{P_2}(G_4 G_5 G_3^\dagger) =
 Z_{P_1 + P_2} (G_1 G_2 G_4 G_5).
 \end{equation}
 This has the unique solution $Z_P(G) = \langle G | e^{ - (A_P/2N) \Delta} | \mathbb I \rangle$, where $A_P$ is the area of the plaquette $P$ and $\Delta$ is
 the Laplacian for the classical group $G$. Moreover, expanding about $U \mapsto \mathbb I + i A_\mu d x^\mu$ shows that in the continuum limit
 $Z_P$ reduces to $e^{-I_{\rm YM}}$.
 
 The property (\ref{ZZ}) is used to argue that the corresponding lattice representation is in fact exact for any triangulation. Using the smallest
 number of triangles consistent with the topology of $\mathcal M$, $Z_{\mathcal M}$ can thus be computed exactly. Consider in particular the
 case that $\mathcal M$ is a sphere of area $A$. One first notes that this quantity together with the rank $N$ of the classical group can
 be taken as the parameters in (\ref{ZG}), allowing the coupling $\lambda^2$ to be scaled out of the problem by setting $\lambda^2 N = 1$.
 Writing then $Z_{\mathcal M}(\lambda;G) = Z(A;G)$, and with
 \begin{equation}\label{Van}
 \Delta(x_1,\dots,x_N) = \prod_{1 \le j < k \le N} (x_k - x_j),
 \end{equation}
 the Vandermonde product, the exact partition functions obtained this way for the group $G$ equal to one of the classical groups are \cite{Ru90,CNS96}
 \begin{align}\label{full}
 Z(A;{\rm U}(N)) \propto e^{- A (N^2 - 1)/24} \sum_{n_1,\dots,n_N = - \infty}^\infty \Delta^2(n_1,\dots,n_N) e^{-(A/2N) \sum_{j=1}^N n_j^2}\nonumber \\
 Z(A;{\rm Sp}(2N)) \propto e^{A(N+1/2)(N+1)/12}  \sum_{n_1,\dots,n_N = - \infty}^\infty \Delta^2(n_1^2,\dots,n_N^2) (\prod_{j=1}^N n_j^2)
e^{-(A/4N) \sum_{j=1}^N n_j^2} \nonumber \\
 Z(A;{\rm O}^+(2N)) \propto e^{A(N-1/2)(N-1)/12}  \sum_{n_1,\dots,n_N = - \infty}^\infty \Delta^2(n_1^2,\dots,n_N^2) 
e^{-(A/4N) \sum_{j=1}^N n_j^2},
\end{align}
where the proportionality constants depend only on $N$. Like the matrix integrals in (\ref{25}), these expressions can be analyzed in the double
scaling limit and give rise to a Painlev\'e II transcendent. In fact the functional forms obtained are identical to those for the corresponding case of
(\ref{25}), as will now be reviewed.

\subsection{The double scaling limit}
Consider for definiteness the group integral in (\ref{25}) in the case $G=U(N)$.
In the limit $N \to \infty$ this exhibits the leading order form \cite{GW80}
\begin{equation}\label{3}
\lim_{N \to \infty} {1 \over N^2} \log  \Big \langle e^{b N {\rm Tr} (U + U^\dagger)} \Big \rangle_{U \in {\rm U}(N)}  =
\left \{
\begin{array}{ll} b^2, & 0 < b < {1 \over 2} \\ \displaystyle
2b - {3 \over 4} - {1 \over 2} \log 2b, & b > {1 \over 2}, \end{array} \right.
\end{equation}
and so is discontinuous in the third derivative at $b=1/2$. The first rigorous proof of this
result was given by Johansson \cite{Jo98}. As mentioned, a string theory viewpoint of two-dimensional QCD motivated tuning the
coupling $b$ to scale with $N$ as it approaches $1/2$.  Specifically, setting $b=1/2 - 2^{-4/3} N^{-2/3} t$ it was shown by Periwal and Shevitz 
\cite{PS90a} using orthogonal polynomial methods that then
\begin{equation}\label{3a}
{d^2 \over d t^2} \lim_{N \to \infty} \Big (  \log  \Big \langle e^{b N {\rm Tr} (U + U^\dagger)} \Big \rangle_{U \in {\rm U}(N)}  - b^2 N^2 \Big ) = - q_0^2(t),
\end{equation}
where $q_0(t)$ satisfies the Painlev\'e II equation (\ref{1.1}) in the case $\alpha = 0$. Of course one still has to determine the boundary condition.
This was done by Crnkovi\'c et al.~\cite{CDM92}, who showed that it is required  $q_0(t) \to (- {1 \over 2} t)^{1/2}$ as $t \to - \infty$ and $q_0(t) \to 0$ as
$t \to \infty$. Moreover, it  was pointed out in \cite{CDM92} that the existence and uniqueness of such a solution of  Painlev\'e II
has been known since the work of Hastings and Macleod \cite{HM80}. In the latter work the condition  $q_0(t) \to 0$ as $t \to \infty$ is further refined to
\begin{equation}\label{HM}
q_0(t) \mathop{\sim}\limits_{t \to \infty} {\rm Ai} (t),
\end{equation}
where Ai$(t)$ denotes the Airy function.

Consider next $Z(A;{\rm U}(N))$ as given in (\ref{full}). It was shown by Douglas and Kazakov \cite{DK93} that
\begin{equation}\label{36}
\lim_{N \to \infty} {1 \over N^2} \log Z(A;{\rm U}(N)) = \left \{  \begin{array}{ll} F_-(A), & A \le \pi^2, \\
F_+(A), & A \ge \pi^2. \end{array} \right.
\end{equation}
The function $F_-$ can be expressed explicitly in terms of elementary functions, while $F_+$ has an explicit form in terms of elliptic
functions. For present purposes, the important feature is that when expanded about $A = \pi^2$ the two functions agree up
to order $(A - \pi^2)^2$, and thus like (\ref{3}) there is a third order phase transition. Gross and Matytsin \cite{GM94} analyzed the corresponding
double scaling limit, obtaining the result
\begin{equation}\label{36a}
 \lim_{N \to \infty}   {d^2 \over dt^2} \log e^{-A^2 N^2/4 \pi^4}  Z(A;{\rm U}(N)) \Big |_{A = \pi^2 - t \pi^2/(2N)^{2/3}} = - q_0^2(t),
\end{equation}
where $q_0(t)$ is the very same Painlev\'e II transcendent as in (\ref{3a}).

The double scaling limit of the group integral in (\ref{25}) in the cases of the orthogonal, or unitary symplectic classical groups was first undertaken by
Myers and Periwal \cite{MP90x}.  They obtained the results
\begin{equation}\label{A1}
  \lim_{N \to \infty}  {d^2 \over d t^2} \Big (  \log  \Big \langle e^{b N {\rm Tr} O} \Big \rangle_{O \in {\rm O}^+(2N)}  - 2 b^2 N^2 \Big )  =
-{1 \over 2} (q_0^2(t) - q_0'(t) ) 
\end{equation}
and 
\begin{align}\label{A2}
 \lim_{N \to \infty}  {d^2 \over d t^2}\Big (  \log  \Big \langle e^{b N {\rm Tr} O} \Big \rangle_{O \in {\rm O}^+(2N+1)}  - 2 b^2 N^2 \Big ) 
& =  \lim_{N \to \infty}  {d^2 \over d t^2}  \Big (  \log  \Big \langle e^{b N {\rm Tr} S} \Big \rangle_{S \in {\rm Sp}(2N)}  - 2 b^2 N^2 \Big ) \nonumber \\
& = -{1 \over 2} (q^2_0(t) + q'_0(t) ),
\end{align}
where $b$ is scaled as specified above (\ref{3a}) but with $N$ replaced by $2N$. We observe that adding (\ref{A1}) and (\ref{A2}) gives
(\ref{3a}). This can be understood as a result of the more general identity \cite{BR01a}
\begin{equation}\label{n.3N}
\Big \langle \prod_{j=1}^{2n+1}  a(e^{i \theta_j})
\Big \rangle_{{\rm U}(2n+1)}  =
\Big \langle \prod_{j=1}^{n} a(e^{i \theta_j})
\Big \rangle_{\widetilde{{\rm Sp}(2n)}}
\Big \langle \prod_{j=1}^{n+1} a(e^{i \theta_j})
\Big \rangle_{\widetilde{{\rm O}^+(2n+2)}}
\end{equation}
where it is required that $a(e^{i \theta}) = a(e^{-i \theta})$, and the tilde symbol  denotes that only eigenvalues $0 < \theta_j < \pi$ are
considered in the average.

It remains to consider the cases of the orthogonal and unitary symplectic classical groups in (\ref{3a}). Their double scaling limit was first evaluated by 
Crescimanno et al.~\cite{CNS96}. To present the results, it is convenient to introduce
\begin{align}\label{EF1}
\tilde{E}_N(L) & \propto {1 \over L^{2N^2 - N}}  \sum_{n_1,\dots,n_N = -\infty}^\infty \Delta^2(n_1^2,\dots,n_N^2) 
e^{-(\pi^2/2L^2) \sum_{j=1}^N n_j^2} \nonumber \\
\tilde{F}_N(L) & \propto {1 \over L^{2N^2 + N}}  \sum_{n_1,\dots,n_N = 1}^\infty \Delta^2(n_1^2,\dots,n_N^2) (\prod_{j=1}^N n_j^2)
e^{-(\pi^2/2L^2) \sum_{j=1}^N n_j^2}
\end{align}
where the proportionality constants depend on $N$ only. Comparing with (\ref{full}) in the case $A = 2 \pi^2/L^2$ shows
\begin{align}\label{EF}
\tilde{E}_N(L) & \propto e^{\pi^2 N(N-1/2)(N-1)/6L^2} L^{-2N^2 + N} Z \Big ( {2 \pi^2 N \over L^2}; {\rm O}^+(2N) \Big )  \nonumber \\
\tilde{F}_N(L) & \propto e^{\pi^2 N(N+1/2)(N+1)/6L^2} L^{-2N^2 - N}
Z \Big ( {2 \pi^2 N \over L^2}; {\rm Sp}(2N) \Big ).
\end{align}
In terms of these quantities, the working of \cite{CNS96}, further refined by calculations in \cite{FMS11} and \cite{SMCF12}, gives that
\begin{align}\label{A3}
\lim_{N \to \infty} {d^2 \over d t^2} \log \tilde{E}_N \Big ( \sqrt{2N}(1 + t/(2^{7/3} N^{2/3})) \Big ) &=  - {1 \over 2} ( q^2_0(t) - q'_0(t) ), \nonumber \\
\lim_{N \to \infty} {d^2 \over d t^2} \log \tilde{F}_N \Big ( \sqrt{2N}(1 + t/(2^{7/3} N^{2/3})) \Big ) & = - {1 \over 2} ( q^2_0(t) + q'_0(t) ).
\end{align}
Thus we have that (\ref{3a}) the double scaling limits (\ref{3a}) and (\ref{36a}) are identical, as are the results (\ref{A1}) and (\ref{A2}) to
(\ref{A3}). Furthermore, as for the group integrals, only two of the three double scaling limits are independent due to the identity
\cite{Fo06}
\begin{eqnarray}
&& \sum_{n_1,\dots,n_{2N} = - \infty}^\infty \prod_{l=1}^{2N} g(n_l) \,
\prod_{1 \le j < k \le 2N} (n_j - n_k)^2  
 \propto
\Big ( \sum_{n_1,\dots,n_{N} = - \infty}^\infty \prod_{l=1}^{N} g(n_l) \,
\prod_{1 \le j < k \le N} (n_j^2 - n_k^2)^2 \Big ) \nonumber \\
&& \hspace*{5cm}  \times 
 \Big ( \sum_{n_1,\dots,n_{N} = - \infty}^\infty \prod_{l=1}^{N} n_l^2 g(n_l) \,
\prod_{1 \le j < k \le N} (n_j^2 - n_k^2)^2 \Big ),
\end{eqnarray}
valid for general $g$ even.

The limit formulas reviewed in this section were derived as formally exact results, but without
rigor. In distinction, rigorous working accompanied the limit formulas obtained in the setting
of the next two subsections.

\subsection{Fluctuation formulas}
We now turn our attention to a seemingly unrelated class of problems, coming  from statistical physics. To introduce these problems, consider a unit square containing
$n$ points with Poisson distribution ${\rm Pr}(n) = {\lambda^{2n} \over n!} e^{-\lambda^2}$.  Starting at $(0,0)$, join dots with line segments of positive slope such
that a continuous path ending at $(1,1)$ is formed, and record the number of dots in the path. Call the maximum number of dots in such a path  $h^\square$. This is a random variable which defines the so called Hammersley process (see e.g.~\cite[\S 10.9]{Fo10}). From work of Gessel \cite{Ge90} and Rains \cite{Ra98}
we have that
\begin{equation}\label{4}
{\rm Pr}(h^\square<N)=e^{-\lambda^2} \Big \langle e^{ \lambda {\rm Tr}(U+U^\dagger)}
\Big \rangle_{U \in {\rm U}(N)}
\end{equation}
(the first of these references actually gives a Toeplitz integral formula, which can be shown to be equivalent to (\ref{4})). This then provides a
probabilistic interpretation of the single plaquette partition function in (\ref{25}) with gauge group equal to ${\rm U}(N)$. A rigorous analysis of 
Baik et al.~\cite{BDJ98} established that
\begin{equation} \label{fiftheqn}
\lim_{\lambda \rightarrow \infty} {\rm Pr} \bigg(\frac{h^\square-2\lambda }{\lambda^{1/3}}<t\bigg)= \exp\bigg(-\int_t^\infty (s-t)q^2_0(s) \,ds\bigg),
\end{equation}
which one sees is completely consistent with (\ref{3a}).

Rains \cite{Ra98} formulated several symmetrized versions of the Hammersely process, in particular by requiring that the points be symmetrical
about the diagonal (line joining (0,0) to (1,1)), or the antidiagonal (line joining (0,1) to (1,0)). Again one wants to record the maximum number of
dots in a directed path from $(0,0)$ to (1,1). Let this number be denoted $h^\symmS$ and $h^\symmO$ respectively. Then we have from
\cite{Ra98} that
\begin{align}\label{R1}
{\rm Pr}(h^\symmS<2N) & = e^{-\lambda^2} \Big \langle e^{ \lambda {\rm Tr} S}
\Big \rangle_{S \in {\rm Sp}(2N)} \nonumber \\
{\rm Pr}(h^\symmO<N) & = {1 \over 2} e^{-\lambda^2} \Big ( \Big \langle e^{ \lambda {\rm Tr} O}
\Big \rangle_{O \in {\rm O}^+(N)}  +
\Big \langle e^{ \lambda {\rm Tr} O}
\Big \rangle_{O \in {\rm O}^-(N)}  \Big ),
\end{align}
so providing a probabilistic interpretation of the single plaquette partition function in (\ref{25}) with gauge group equal to ${\rm Sp}(2N)$ or
${\rm O}(N)$.
Moreover, rigorous working in \cite{BR01a} established that
\begin{align}\label{R2}
\lim_{\lambda \rightarrow \infty} {\rm Pr} \bigg(\frac{h^\symmS-2\lambda }{\lambda^{1/3}}<t \bigg) &= \exp\bigg(-{1 \over 2} \int_t^\infty (x-t)q^2_0(x) \, dx + {1 \over 2}
\int_t^\infty q_0(x) \, dx
\bigg), \nonumber \\
\lim_{\lambda \rightarrow \infty} {\rm Pr} \bigg(\frac{h^\symmO-2\lambda }{\lambda^{1/3}}<t \bigg) &= {1 \over 2}
\bigg \{ \exp\Big (-{1 \over 2} \int_t^\infty (x-t)q^2_0(x) \, dx  + { 1 \over 2}
\int_t^\infty q_0(x) \, dx  \Big ) \nonumber \\
& +  \exp\Big (-{1 \over 2} \int_t^\infty (x-t)q^2_0(x) \, dx - {1 \over 2}
\int_t^\infty q_0(x) \, dx,
\Big) \bigg \}, 
\end{align}
which, like the relation between (\ref{fiftheqn}) and (\ref{3a}), is completely consistent with (\ref{A3}).

It is furthermore the case that the Yang-Mills partition functions (\ref{full}) have interpretations in statistical physics
\cite{FMS11}, \cite{SMCF12}. The most significant arises in the case of the gauge group Sp$(2N)$. A Brownian motion 
confined to the half-line $x \ge 0$ and conditioned to start at $x=0$ when $t=0$ and to finish at $x=1$  when $t=1$ is referred to
as a Brownian excursion. Suppose $N$ such Brownian motions are placed at $x=0$ when $t=0$, and furthermore are conditioned
to not intersect for $0<t<1$, before all returning to $x=0$ when $t=1$. Let $x_N(t)$ denote the position of the rightmost Brownian motion in
$[0,1]$, and define the random variable $H_N = {\rm max} \, \{x_N(t), \: 0 < t < 1 \}$ specifying the maximum displacement. It was shown
in \cite{SMCR08,KIK08,Fe12} that
\begin{equation}
{\rm Pr} \, (H_N \le L) =  \tilde{F}_N (L),
\end{equation}
where $ \tilde{F}_N (L)$ is specified by (\ref{EF1}) with the particular proportionality constant
$$
A_N = {\pi^{2 N^2 + N} \over 2^{N^2 + N/2} \prod_{j=0}^{N-1} \Gamma(2 + j) \Gamma(3/2 + j)}.
$$
Using the orthogonal polynomial method of \cite{GM94} and \cite{CNS96}, it was shown by Forrester et al.~\cite{FMS11} that
\begin{equation}
\lim_{N \to \infty} \tilde{F}_N ( \sqrt{2N} + 2^{-11/6} N^{-1/6} s) = \exp\bigg(-{1 \over 2} \int_s^\infty (x-s)q_0^2(x) \, dx + {1 \over 2}
\int_s^\infty q_0(x) \, dx
\bigg).
\end{equation}
Subsequently Liechty \cite{Li12} used a Riemann-Hilbert approach to the orthogonal polynomial method to make this rigorous
(see too the related work \cite{CQR11}).

The growth models reviewed in this section all have the feature that they give rise to expressions that appeared earlier in two-dimensional QCD. On should remark that there is now a vast literature on exact fluctuation formulas in growth models, particularly due to their relevance to the Kardar-Parisi-Zhang (KPZ) universality class (see \cite{Co12} for a recent review).

\subsection{Soft edge random matrix distributions}
The distribution functions exhibited in (\ref{R1}) and (\ref{R2}) occur in a probabilistic setting again seemingly unrelated to what
has been discussed to date, namely as the  distribution of the extreme eigenvalue of certain universality classes of
random matrices in the neighbourhood of a soft edge. By way of background, we remark that an Hermitian
 random matrix ensemble is said
to have orthogonal ($\beta = 1$), unitary ($\beta = 2$) and unitary symplectic ($\beta = 4$) symmetry if it has an eigenvalue probability
density function of the form
\begin{equation}\label{Cg}
{1 \over C_N} \prod_{l=1}^N g(\lambda_l) \prod_{1 \le j < k \le N} | \lambda_k - \lambda_j |^\beta.
\end{equation}
The soft edge refers to a spectrum edge such that the leading order eigenvalue density decays to zero like a square root. As a concrete
example, consider the case $g(x) = e^{-\beta x^2/2}$. Then (\ref{Cg}) can be realized by Hermitian matrices with real ($\beta = 1$),
complex ($\beta = 2$) and real quaternion ($\beta = 4$) entries (see e.g.~\cite[Ch.~1]{Fo10}). The entries on and above the diagonal are independent
Guassians of mean zero and appropriate variance (which is different for the diagonal and off diagonal entries). To leading order the eigenvalue density is
supported on $[- \sqrt{2N}, \sqrt{2N}]$, and vanishes like a square root at either end. Moreover, in the neighbourhood of such soft edges a well
defined statistical state results by introducing the scaled variable $s$ according to \cite{Fo93a} $\lambda = \sqrt{2N} + s/(2^{1/2} N^{1/6})$. In particular
the largest eigenvalue has a well defined limiting distribution in terms of these variables. Explicitly, let $E_{\beta,N}(0;(a,b);g(\lambda))$ denote the
probability that for the ensemble specified by (\ref{Cg}) that are no eigenvalues in the interval $(a,b)$. Then the distribution of the largest eigenvalue
at the soft edge is specified by the limit
\begin{equation}\label{Cg1}
E_\beta^{\rm soft}(0;(s,\infty)) := \lim_{N \to \infty}  E_{\beta,N}(0;( \sqrt{2N} + s/(2^{1/2} N^{1/6}),\infty);e^{-\beta \lambda^2/2}).
\end{equation}

Beginning with a Fredholm determinant formula for (\ref{Cg1}) in the case $\beta = 2$  \cite{Fo93a}, Tracy and Widom \cite{TW94a} deduced that
\begin{equation}\label{Cg2}
E_2^{\rm soft}(0;(s,\infty)) =  \exp\bigg(-\int_s^\infty (x-s)q^2_0(x) \,dx\bigg)
\end{equation}
(see \cite{ASV95,FW00,BD00} for subsequent derivations).
This result was known before  (\ref{fiftheqn}), thus allowing for the interpretation of the latter in terms of the scaled largest eigenvalue of a complex
Hermitian random matrix.  Subsequently Tracy and Widom \cite{TW96} gave Painlev\'e formulas for the evaluation of  the limit (\ref{Cg1}) in the
cases $\beta = 1$ and 4. The former of these reads
\begin{equation}\label{Cg3}
E_1^{\rm soft}(0;(s,\infty)) =  \exp\bigg(-{1 \over 2} \int_s^\infty (x-s)q^2_0(x) \, dx + {1 \over 2}
\int_s^\infty q_0(x) \, dx
\bigg).
\end{equation}
It was pointed out by Forrester and Rains \cite{FR01} that the case $\beta = 4$ can be deduced from knowledge of the $\beta = 1$ and 2 results,
and furthermore is most naturally expressed in terms of the rescaled quantity
\begin{equation}\label{Cg4}
\tilde{E}_4^{\rm soft}(0;(s,\infty)) = \lim_{N \to \infty}  E_{\beta,N/2}(0;( \sqrt{2N} + s/(2^{1/2} N^{1/6}),\infty);e^{- \lambda^2/2}),
\end{equation}
when it reads
\begin{align}\label{Cg5}
\tilde{E}_4^{\rm soft}(0;(s,\infty)) & =  {1 \over 2}
\bigg \{  \exp\Big (-{1 \over 2} \int_s^\infty (x-s)q^2_0(x) \, dx +  {1 \over 2}
\int_s^\infty q_0(x) \, dx  \Big )\nonumber \\
&+  \exp\Big (-{1 \over 2} \int_s^\infty (x-s)q^2_0(x) \, dx- {1 \over 2}
\int_s^\infty q_0(x) \, dx
\Big) \bigg \}.
\end{align}
Again, the evaluations (\ref{Cg3}) and (\ref{Cg5})  allowed the results (\ref{R2}) to be interpreted in terms of the scaled largest eigenvalue of a real, and real quaternion respectively,
Hermitian random matrix. 

We remark that the matrix integrals in (\ref{25}) for each of the classical groups satisfy transformation identities relating them to the gap probabilities
$E_{\beta}^{\rm hard}(0;(0,s^2);a)$ with $a=N$ and $s$ proportional to $bN$ \cite{BF03}. 
The general $\beta$ hard edge state with parameter $a$ is realized by choosing $g(\lambda) = \lambda^a e^{-\beta \lambda/2}$, $\lambda > 0$ in
(\ref{Cg}), then scaling all the eigenvalues $\lambda_l \mapsto x_l/(4N)$ and taking the limit $N \to \infty$. It is called a hard edge state because
the eigenvalue density is strictly zero for $\lambda < 0$, and diverges like $\lambda^{-1/2}$ as $\lambda \to 0^+$. As $a \to \infty$ there is a hard-to-soft
edge transition which gives an explanation for (\ref{fiftheqn}) and (\ref{R2}).

A Painlev\'e evaluation of $E_2^{\rm soft}(0;(s,\infty))$ seemingly different to (\ref{Cg2}) was given by Forrester and Witte \cite{FW00}. For this, introduce
the Jimbo-Miwa-Okamoto $\sigma$-form of the Painlev\'e II equation
\begin{equation}\label{JM}
(u_a'')^2 + 4 u_a'( (u_a')^2 - t u_a' + u_a) - a^2 = 0.
\end{equation}
Then it was shown  that
\begin{align}\label{Cg6}
E_2^{\rm soft}(0;(s,\infty)) & = \exp \Big ( - \int_s^\infty u_0(x) \, dx \Big ) \nonumber \\
& =   \exp \Big ( - \int_s^\infty (s-x) u_0'(x) \, dx \Big ),
\end{align}
where $u_0(x)$ satisfies (\ref{JM}) with $a=0$ subject to the boundary condition
\begin{equation}\label{tA}
u_0(x) \mathop{\sim}\limits_{x \to \infty} ({\rm Ai}'(x))^2 - x ({\rm Ai}(x))^2. 
\end{equation}
For future reference we remark that (\ref{tA}) implies that $u_0'(x)$ has the asymptotic behaviour
\begin{equation}\label{tA1x}
u_0'(x) \mathop{\sim}\limits_{x \to \infty} -({\rm Ai}(x))^2.
\end{equation}

Theory relating to (\ref{JM}) \cite{Ok86} tells us that
\begin{equation}\label{tA1}
u_0'(t) = - 2^{-1/3} \Big ( q_{1/2}'(t) + q_{1/2}^2(t) + {t \over 2} \Big ) \Big |_{t \mapsto - 2^{1/3} t}
\end{equation}
(see in particular \cite[eq.~(5.27)]{FW00}; note that in this work (\ref{1.1}) is written with
$\alpha \mapsto - \alpha$)
where $q_\alpha$ denotes a solution of the Painlev\'e equation (\ref{1.1}), and so we have that $E_2^{\rm soft}(0;(s,\infty))$ can be
evaluated in terms of $q_0$ as specified by (\ref{Cg2}), or in terms of $q_{1/2}$ as specified by (\ref{Cg6}) and (\ref{tA1}).
As noted in \cite{FW00}, these two forms can be reconciled by invoking an identity of Gambier \cite{Ga09} which gives
\begin{equation}\label{tA1a}
 \epsilon 2^{1/3} q_0^2(-2^{-1/3} t) = {d \over dt} q_{\epsilon/2}(t) + \epsilon q_{\epsilon/2}^2(t) + {\epsilon \over 2} t,
\end{equation}
valid for both $\epsilon = \pm 1$, and thus in particular $u'(t;0) = -q_0^2(t)$.

\subsection{Parameter dependent problems and Lax pairs}
Underlying the analysis relating the Hammersely process to random matrix averages is a bijection between permutations and standard
tableaux due to Robinson and Schensted (see e.g.~\cite{Fu97}). As also detailed in \cite{Fu97}, this was generalized by Knuth to give a
bijection between weighted non-negative integer matrices and  weighted semi-standard tableaux (or equivalently non-intersecting lattice paths).
The latter underlies a class of lattice extensions of the Hammersely model introduced by Johansson \cite{Jo99a} and further generalized by
Baik and Rains \cite{BR01}. These are defined by an $n_1 \times n_2$ rectangular grid of lattice points $(i,j)$, each specifying a non-negative
random variable $x_{i,j}$. The primary quantity of interest is the distribution of the maximum of the random variables when summed
over an up/right path starting at the bottom left and finishing at the top right, referred to as the last passage time (this has the alternative
interpretation as the maximum displacement in the non-intersecting lattice paths picture).

In the case of a $n \times (n+1)$ grid, $x_{i,j}$ ($j \ne n +1$) Poisson distributed with unit mean, $x_{i,n+1}$ Poisson distributed with mean
$(1+a_1)/2$, the last passage time is distributed as for $x_1$ in the probability density function proportional to
\begin{equation}\label{B1}
e^{- \sum_{j=1}^n (x_j + y_j)/2 - a_1 \sum_{j=1}^n (x_j - y_j)/2}
\prod_{1 \le i < j \le n} (x_i - x_j) ( y_i - y_j)
\end{equation}
where $x_1 > y_1 > \cdots > x_n > y_n > 0$. It was proved in \cite{BR01a} that
\begin{equation}\label{2.A}
\lim_{n \to \infty} {\rm Pr} \, (x_1 < 4n + 2 (2n)^{1/3} s ) \Big |_{a_1 = - w/(2 (2n)^{1/3}) }=
f(s;w) E_2^{\rm soft}(0;(s,\infty)),
\end{equation}
where $f$ relates to the Painlev\'e II transcendent $q_0(s)$ via the Lax pair equations
\begin{align}\label{Aa}
{\partial \over \partial s} \begin{bmatrix} f(s;w) \\ g(s;w) \end{bmatrix} & =
\begin{bmatrix} 0 & q_0(s) \\
q_0(s) & - w \end{bmatrix}  \begin{bmatrix} f(s;w) \\ g(s;w) \end{bmatrix} \nonumber \\
{\partial \over \partial w }  \begin{bmatrix} f(s;w) \\ g(s;w) \end{bmatrix}  & =
\begin{bmatrix} (q_0(s))^2  &  -w q_0(s) - q_0'(s) \\
-w q_0(s) + q_0'(s)&  w^2 - s - (q_0(s))^2 \end{bmatrix}  \begin{bmatrix} f(s;w) \\ g(s;w) \end{bmatrix}. 
\end{align}
(By definition of a Lax pair, $q_0(s)$ as it appear is (\ref{Aa}) is determined by consistency of the mixed derivatives
${\partial^2 \over \partial s \partial w}$ and ${\partial^2 \over \partial w \partial s}$, and this gives rise to (\ref{1.1}) with
$\alpha = 0$.)

It is also the case that both $f$ and $g$ from the Lax pair appear as the distribution function for a particular symmetrized version of the
above last passage time model \cite{BR01a, BR01}. This is specified by a $2n \times 2n$ grid with $x_{i,j}$ $(i < j)$ Poisson distributed with unit mean,
$x_{i,j} = x_{j,i}$ for $i > j$ and $x_{i,i}$ Poisson distributed with mean $(1 + a_1)/2$. The last passage time is distributed as for $x_1$ in the probability
density function proportional to
\begin{equation}\label{B2}
e^{- \sum_{j=1}^{2n} x_j/2} e^{- a_1 \sum_{j=1}^{2n} (-1)^{j-1} x_j/2} \prod_{1 \le i < j \le 2n} (x_j - x_i),
\end{equation}
where $x_1 > x_2 > \cdots > x_{2n} > 0$, and it was proved in \cite{BR01a} that the distribution function permits the scaling limit
\begin{eqnarray}\label{2.B}
\lefteqn{ \lim_{n \to \infty} {\rm Pr} \, ( x_1 < 4n + 2(2n)^{1/3} s ) \Big |_{a_1 = - \alpha/(2 (2n)^{1/3} )} }\nonumber \\
\quad  &&= {1 \over 2} \Big ( (f + g) e^{ {1 \over 2} \int_s^\infty q_0(t) \, dt} + (f - g)   e^{ -{1 \over 2} \int_s^\infty q_0(t) \, dt} \Big )
\Big ( E_2^{\rm soft}(0;(s;(s,\infty))) \Big )^{1/2}.
\end{eqnarray}

A random matrix interpretation of (\ref{B1}) and (\ref{B2}), and thus of (\ref{2.A}) and (\ref{2.B}), was provided in \cite{FR02b}.
Specifically, we have that (\ref{B1}) with $a_1 = 2/b - 1$ is the eigenvalue probability density function of the rank 1 perturbed
complex Wishart matrix, 
\begin{equation}\label{M}
M = X^\dagger X + b \vec{x} \vec{x}^\dagger, 
\end{equation}
where $X$ ($\vec{x}$) is an $n \times n$ ($n \times 1$) matrix with
standard complex Gaussian entries. With regards to (\ref{B2}), let $X$ in (\ref{M}) be a $2n \times 2n$ antisymmetric standard complex Gaussian
matrix, and let $\vec{x}$ be a $2n \times 1$ standard complex Gaussian vector. Then we have that the perturbed eigenvalues of $M$ have PDF
(\ref{B2}) with $a_1 = 2/b - 1$.

It has been known since the work of Jones et al.~\cite{JKT78} that rank 1 perturbed Gaussian ensembles can exhibit a separation of the
largest eigenvalue when the strength of the coupling exceeds a critical value. The significance of the coupling implied by (\ref{2.A}) and
(\ref{2.B}) is that it corresponds to a scaling about this critical value, while the scaling of $x_1$ centres about the leading order value of
the largest eigenvalue, and measures in units such that the largest eigenvalues have spacing of order unity. The description of the rank 1
perturbed scaled soft edge state for $\beta$-ensembles ((\ref{2.A}) corresponds to $\beta = 2$; (\ref{2.B}) to $\beta = 4$) has been the subject
of a number of recent papers \cite{BV11,Wa11,Fo11y, Mo11}.

Another setting which gives rise to a Lax pair for $q_0(s)$ is the analogue of the double scaling limit for the so-called Hermitian quartic matrix
model ((\ref{Cg}) in the case $\beta = 2$, $g(\lambda) = e^{-(t/2) \lambda^2 + (g/4) \lambda^4}$ with $(1 - t^2/(4g)) N^{2/3} = O(1)$)
\cite{BI03}. So
does the scaled distribution of the maximum height in the nonintersecting Brownian excursion model detailed below (\ref{R2}) when considered also
as a function of the scaled location of this maximum about $t=1/2$ \cite{Sc12,QR12,BLS12}.

We conclude this section by summarising a very recent calculation \cite{WBF12} of the joint distribution $p_{2,(2)}^{\rm soft}(x_1,x_2)$,
of the scaled largest ($x_1$) and second largest ($x_2$)  eigenvalue
at the soft edge for $\beta = 2$. To present this result, let $p_{2,(1)}^{\rm soft}(t)$ denote the distribution of the scaled largest eigenvalue, so that
$$
p_{2,(1)}^{\rm soft}(t) = {d \over dt} E_2^{\rm soft}(0;(t,\infty)).
$$
Also, with $q_\alpha$ a solution of (\ref{1.1}), let  
\begin{equation}\label{pq}
p_\alpha = q_\alpha' + q_\alpha^2 + {1 \over 2} t, 
\end{equation}
and introduce $U(x,t)$ and $V(x,t)$
through the Lax pair equations
\begin{align}\label{Lax}
& {\partial \over \partial x} \begin{bmatrix} U \\ V  \end{bmatrix} = \nonumber \\
& \: \: =
\Bigg ( \begin{bmatrix} 0 & 0 \\ - {1 \over 2} & 0 \end{bmatrix} x +
 \begin{bmatrix} \displaystyle - q_{3/2} - {2 \over p_{3/2}} & - 1 \\
 \displaystyle  {1 \over 2} (t - p_{3/2}) +  ( q_{3/2} + {2 \over p_{3/2}})^2 &  \displaystyle q_{3/2} + {2 \over p_{3/2}} \end{bmatrix} +
 \begin{bmatrix}  1 & p_{3/2} \\ 0 & -1 \end{bmatrix} {1 \over x}   \Bigg )  \begin{bmatrix} U \\ V  \end{bmatrix}, \nonumber \\
 &  {\partial \over \partial t} \begin{bmatrix} U \\ V  \end{bmatrix}  =
\Bigg (   \begin{bmatrix} 0 & 0 \\ {1 \over 2} & 0 \end{bmatrix} x + 
  \begin{bmatrix} 0 & 1 \\  0  &   \displaystyle  -2 ( q_{3/2} + {2 \over p_{3/2}} ) \end{bmatrix}  \Bigg )
 \begin{bmatrix} U \\ V  \end{bmatrix}
  \end{align}  
(it is noted in  \cite[Remark 1]{WBF12}  that (\ref{Lax}) are a transformed version of a Lax pair for PII first given by Flashka and Newell \cite{FN80}; furthermore, as in \cite{FW00}, (\ref{1.1}) is
written with $\alpha \mapsto - \alpha$).
With this notation, we have from  \cite[Prop.~9 and 10]{WBF12} that, subject to some specific boundary conditions for the transcendents involved,
\begin{align}
p_{2,(2)}^{\rm soft}(t,t-x)  = & {1 \over 4 \pi}  p_{2,(1)}^{\rm soft}(t) t^{-5/2} e^{- {4 \over 3} t^{3/2}} (U \partial_x V - V \partial_x U) ( - 2^{1/3} x; - 2^{1/3} t)\nonumber \\
&
\times \exp \Big ( \int_{2^{1/3} t}^\infty dy \,
\Big ( (2 q_{3/2} +  {4 \over p_{3/2} })(-y) - \sqrt{2y} - {5 \over 2 y}  \Big ) \Big ). 
\end{align}
In the Appendix of \cite{WBF12} it is shown how $q_{3/2}$, through a combination of the (inverse) Gambier identity, 
\begin{equation}\label{2.47}
q_{\epsilon/2}(-2^{1/3}s;\xi) = - \epsilon 2^{-1/3} {d \over ds} \log q_0(s;\xi), \quad \epsilon = \pm 1,
\end{equation}
and a Schlesinger transformation,
can be related to the Hasting-Macleod solution $q_0$ (see also Section \ref{S32} below).

\section{Painlev\'e II with general $\alpha$ in random matrix theory}
\setcounter{equation}{0}
\subsection{Two-dimensional QCD with quarks and matrix models}
In 1991 Minnahan \cite{Mi91} pointed out that the generalization of the one plaquette partition function (\ref{25}) with $G={\rm U}(N)$ 
for two-dimensional QCD
without quarks to the case that each plaquette contains $M$ massless quark flavours is given by 
\begin{equation}\label{F1}
\int dU \, d \psi d \chi \, \exp \Big ( b N {\rm Tr} \, (U + U^\dagger) + \sum_{i=1}^M \vec{\chi}_i^\dagger ( \mathbb{I} - U) \vec{\psi}_i +
 \vec{\psi}_i^\dagger  ( \mathbb{I} - U^\dagger)  \vec{\chi}_i \Big ),
 \end{equation}
 where $\vec{\psi}_i$ and $\vec{\chi}_i$ and $N \times 1$ vectors with components which either commute (bosonic case, $\epsilon = 1$),
 or anticommute (fermionic case, $\epsilon = -1$). Moreover, it is noted that after integrating out over $\psi$ and $\chi$, (\ref{F1}) reduces to
 \begin{equation}\label{F2}
 \Big \langle \exp \Big ( b N {\rm Tr} \, (U + U^\dagger) \mp 2 \epsilon M \log | \det ( \mathbb{I} - U) | \Big ) \Big \rangle_{U \in U(N)}
 \end{equation}
 (of course as written (\ref{F2}) is singular in the bosonic case and requires regularization).
 
 The quantity of interest is again the limiting ``specific heat'' as on the LHS of (\ref{3a}). The method of orthogonal polynomials is
 applied in \cite{Mi91} to deduce that this evaluates in terms of a general Painl\'eve II transcendent to give $-q_\alpha^2(s)$.
 It was stated that $\alpha$ is proportional to $-M$, although the proportionality was not determined.
 
 In 1998 Akemann et al.~\cite{ADMN98}  studied  the correlation kernel for the random matrix ensemble specified by
 (\ref{Cg}) is the case $g(\lambda) = |\lambda|^a e^{-N V(\lambda)}$, where $V(\lambda)$ is such that the leading
 order eigenvalue density vanishes like a quadratic at $\lambda = 0$ (see \cite[\S 13.2]{PS11} for a text book treatment of
 the corresponding specific form of $V$). The correlation kernel (see (\ref{5.1}) below, or more explicitly the orthonormal function entering
 therein) was shown to be given as the solution of a differential equation in which the coefficients are solutions of
 the general Painlev\'e II transcendent with $\alpha$ proportional to $a$ (the form of the general Painlev\'e II equation in
 \cite[eq.~(2.50)]{ADMN98} is not in the canonical form (\ref{1.1}), but doing so would give the explicit proportionality). Later
 Claeys et al.~\cite{CKV08} gave a rigorous treatment of this problem using Riemann-Hilbert methods, and expressed the limiting correlation kernel in terms of the
 two-components of the Lax pair formulation of general $\alpha$ Painlev\'e II due to Flascha and Newell \cite{FN80}
 (this also came after the rigorous work of Bleher and Its \cite{BI03} using Riemann-Hilbert methods to treat the
  case $\alpha = 0$).  Moreover, the
  Painlev\'e II  transcendent $q_\alpha(s)$ appearing therein was specified to have the asymptotics
  \begin{equation}\label{HS1}
 q_\alpha(s) \mathop{\sim}\limits_{s \to - \infty}  \sqrt{-s \over 2} \quad {\rm and} \quad
 q_\alpha(s) \mathop{\sim}\limits_{s \to \infty} {\alpha \over s}
 \end{equation}
and this solution was proved to be
 free of poles on the real axis, thus generalizing properties of the Hasting-Macleod $\alpha = 0$ solution appearing in Section 2.
 
 \subsection{Random matrix averages}\label{S32}
 The eigenvalue PDF for the Gaussian unitary ensemble (GUE) corresponds to the case $g(\lambda) = e^{- \lambda^2}$ of
(\ref{Cg}). Let $\xi^* := 1 - e^{\pi i \mu} (1 - \xi)$. The study of the averages
\begin{eqnarray}\label{A}
\tilde{E}_{\beta,N}(\lambda;\mu;\xi) & := \Big \langle \prod_{l=1}^N (1 - \xi \chi_{(-\infty,\lambda)} )(\lambda - \lambda_l )^\mu \Big \rangle, \nonumber \\
& =  \Big \langle \prod_{l=1}^N (1 - \xi^* \chi_{(-\infty,\lambda)}) |\lambda - \lambda_l |^\mu \Big \rangle
\end{eqnarray}
where $\chi_J^{(l)} = 1$ for $\lambda_l \in J$, $\chi_J^{(l)} = 0$ otherwise was undertaken by Forrester and Witte \cite{FW00,FW03a}
(for a later treatment, see \cite{CF06}). Moreover, the corresponding soft edge scaled quantity
$$
{d \over ds} \log E_2^{\rm soft}(s;\mu;\xi) :=
{d \over ds} \log \Big ( e^{-\mu \lambda^2/2} \tilde{E}_N(\lambda;\mu;\xi) \Big ) \Big |_{\lambda \mapsto \sqrt{2N} + s/\sqrt{2} N^{1/6}}
$$
was computed in \cite{FW01} (see also \cite[Ch.~8]{Fo10}) with the result
\begin{equation}\label{up}
{d \over ds} \log E_2^{\rm soft}(s;\mu;\xi) = u_\mu(s;\xi),
\end{equation}
where $u_\mu$ satisfies (\ref{JM}) with $a=\mu$.

From the definitions
$$
E_2^{\rm soft}(s;\mu=0;\xi=1) = E_2^{\rm soft}(0;(s,\infty)),
$$
where $E_2^{\rm soft}(0;(s,\infty))$ is given by (\ref{Cg1}), and that $u_0(s;\xi=1) = u(s)$. For $\mu = 0$ and general $\xi$, by
expressing $E_2^{\rm soft}(s;\mu=0;\xi)$ as a Fredholm determinant \cite{Fo93a} (see Section \ref{S33} below)
it follows that the boundary condition (\ref{tA}) is to be extended
by  multiplying by $\xi$,
\begin{equation}\label{ud}
u_0(x;\xi) \mathop{\sim}\limits_{x \to \infty} \xi ( ({\rm Ai}'(x))^2 - x ({\rm Ai}(x))^2) \sim {\xi \over 8 \pi x} e^{- {4 \over 3} x^{3/2}}.
\end{equation}
Furthermore, as a consequence of the Gambier identity (\ref{tA1a}), and (\ref{tA1x}), we have
\begin{equation}\label{3.5a}
u'_0(t;\xi) = - q_0^2(t;\xi),
\end{equation}
where $q_0(t;\xi)$ satisfies (\ref{1.1}) with $\alpha = 0$ subject to the boundary condition
\begin{equation}\label{N1}
q_0(t;\xi)  \mathop{\sim}\limits_{t \to \infty} \sqrt{\xi} {\rm Ai}(t)  \mathop{\sim}\limits_{t \to \infty} 
 \sqrt{\xi} {e^{-(2/3) t^{3/2}} \over 2 \sqrt{\pi} t^{1/4}}.
\end{equation}
We remark that the quantity $E_2^{\rm soft}(s;\mu=0;\xi)$, $0 < \xi \le 1$,  has the probabilistic interpretation  that in
the $\beta = 2$ soft edge state, with each of the eigenvalues deleted with probability $(1 - \xi)$, there are no eigenvalues in
$(s,\infty)$ \cite{BP04}.

Note that (\ref{N1}) with $\xi = 1$ is consistent with (\ref{HM}). In this case we know that \cite{HM80}
\begin{equation}\label{B1a}
q_0(t;\xi=1)   \mathop{\sim}\limits_{t \to- \infty}
\sqrt{-t \over 2}, 
\end{equation}
and this used in (\ref{Cg2}) has the consequence that $E_2^{\rm soft}(0;(s,\infty)) \mathop{\sim}\limits_{s \to -\infty}
e^{- (-s)^3/12}$ \cite{TW94a}. On the other hand, it is known \cite{AS77} that for $0 < \xi < 1$,
$q_0(t;\xi)$ decays to zero as ${t \to -\infty}$ and oscillates, 
\begin{equation}\label{N2}
q_0(t;\xi)  \mathop{\sim}\limits_{t \to - \infty} d (-t)^{-1/4}
\sin \Big ( {2 \over 3} (-t)^{3/2} - {3 \over 4} d^2 \log (-t) - c \Big )
\end{equation}
where 
$$
d^2 = - {1 \over \pi} \log (1 - \xi), \quad
c = {3 \over 2} d^2 \log 2 + {\rm arg} \, \Gamma(1 - {1 \over 2} i d^2) - {\pi \over 4},
$$
and this implies the very different asymptotic behaviour \cite{BP04}, \cite{FS12}
$$
E_2^{\rm soft}(s;\mu=0;\xi ) \mathop{\sim}\limits_{s \to -\infty} \exp(  \log (1 - \xi) (2/3 \pi) (-s)^{3/2}).
$$

We seek the boundary conditions relating to (\ref{up}) for general $\xi$ and values of
$\mu$ different to zero. In the cases $\mu = 1$ and $\mu = 2$ this can be done by
using known formulas which give relations with the case $\mu = 0$. In particular we require the
(inverse) Gambier identity (\ref{2.47}), together with the recurrence
\cite[Eqns.~(3.23) and (5.9)]{FW00}
\begin{equation}\label{b1}
u_{\mu+1}(x;\xi) - u_\mu(x;\xi) = - 2^{1/3} q_{\mu+1/2}(-2^{1/3} x;\xi),
\end{equation}
and the identity \cite[Eq.~(5.22)]{FW00}
\begin{equation}\label{b2}
u_2(x;\xi) = {d \over dx} \log u_0(x;\xi) + u_0(x;\xi).
\end{equation}

\begin{prop}
We have
\begin{equation}\label{b3}
u_1(x;\xi) \mathop{\sim}\limits_{x \to \infty}   - x^{1/2} - {1 \over 4 x} + {5 \over 32 x^{5/2}} + 
O\Big ( {1 \over x^4} \Big )  - \xi {e^{-(4/3) x^{3/2}} \over 64 \pi x^{5/2}}\Big ( 1 + O\Big ( {1 \over x^{3/2}}
\Big )  \Big ),
\end{equation}
and
\begin{equation}\label{b31}
u_2(x;\xi) \mathop{\sim}\limits_{x \to \infty}   - 2x^{1/2} - {1 \over  x} + {17 \over 16 x^{5/2}} + 
O\Big ( {1 \over x^4} \Big )  + \xi {e^{-(4/3) x^{3/2}} \over 256 \pi x^{4}}\Big ( 1 + O\Big ( {1 \over x^{3/2}}
\Big )  \Big ).
\end{equation}
\end{prop}

\noindent
Proof. \quad We set $\mu = 0$ in (\ref{b1}), then substitute for $q_{1/2}(-2^{1/3}x;\xi)$ using
(\ref{2.47}) with $\epsilon = 1$ to deduce
\begin{equation}\label{ux1}
u_1(x;\xi) = u_0(x;\xi)  + {d \over ds} \log q_0(s;\xi).
\end{equation}
Next, we extend (\ref{ud}) by writing
\begin{align}\label{dd0}
 u_0(x;\xi) = & {\xi \over 8 \pi x} e^{- {4 \over 3} x^{3/2}}\Big ( 1 + {c \over 2 x^{3/2} }+
 {c_1 \over 2 x^3} + O\Big ( {1 \over x^{9/2}} \Big ) \Big )  \nonumber \\
 & + {\xi^2 \over 128 \pi^2 x^{5/2}} e^{- {8 \over 3} x^{3/2}}\Big ( 1 + {d\over  x^{3/2}} +
 {d_1 \over  x^3} + O\Big ( {1 \over x^{9/2}} \Big ) \Big ) + O\Big ( \xi^3 e^{- 4 x^{3/2}}\Big ).
 \end{align}
 Substituting for $\sigma$ in  (\ref{JM}) with $a=0$ shows
 \begin{equation}\label{cd1}
 c= - {17 \over 12}, \quad c_1 = \Big ( {35 \over 24} \Big )^2, \quad d = - {13 \over 6}, \quad
 d_1 = {1531 \over 288}.
 \end{equation}
Furthermore, we know from \cite{PS00} in the case $\xi = 1$, and \cite[Prop.~9.4.4] {Fo10}) for general $\xi > 0$ that the
analogous extension of (\ref{N1}) is
\begin{equation}\label{N1z}
q_0(s;\xi) \mathop{\sim}\limits_{s \to \infty}
\sqrt{\xi} {e^{-(2/3) s^{3/2}} \over 2 \sqrt{\pi} s^{1/4}}
\Big \{ 1 -{ \alpha_1 \over {2 \over 3} s^{3/2} } + O \Big ( {1 \over s^3} \Big ) +
{\xi  e^{-(4/3) s^{3/2}}  \over 16 \pi s^{3/2}}
\Big ( 1 - { a_1 \over {2 \over 3} s^{3/2} } + O \Big ( {1 \over s^3} \Big )  \Big ) \Big \},
\end{equation}
where
$$
\alpha_1 = {5 \over 72}, \qquad a_1 = {23 \over 24}.
$$
Substituting (\ref{N1z}) and (\ref{dd0}) in (\ref{ux1}) gives, after straightforward manipulation, the
expansion (\ref{b3}).

To obtain the expansion (\ref{b31}), we only require (\ref{dd0}). Thus we substitute in (\ref{b2})
and expand the logarithm to obtain the stated result. \hfill $\square$

\medskip

The structure exhibited in (\ref{b3}) and (\ref{b31}), where the dependence on the parameter
$\xi$, which itself is not a parameter in the Painlev\'e II equation (\ref{1.1}), occurs only in 
exponentially terms is well known in the theory of Painlev\'e equations (see e.g.~\cite{FIKN06})).
The full asymptotic series of such terms is said to form a trans-series.

The leading $x \to \infty$ asymptotic expansion of $u_\mu(x;\xi)$ for $\mu = \pm 1$ and
$0 < \xi \le 1$ can also be determined. In addition to (\ref{b1}), this requires the expression for
$u(x;\mu;\xi)$ (which is essentially the Hamiltonian for the PII system in  Okamoto's
theory \cite{Ok86}) in terms of $q_\mu(x;\xi)$ \cite[Eq.~(5.26)]{FW00}
\begin{equation}\label{B3}
-2^{-1/3} u_\mu(- 2^{-1/3}t;\xi) = {1 \over 2} \Big ( q'_{1/2 - \mu }(t;\xi) \Big )^2 -
{1 \over 2}  \Big ( q^2_{ 1/2 - \mu }(t;\xi) + {t \over 2} \Big )^2 - \mu q_{ 1/2 - \mu }(t;\xi).
\end{equation}

\begin{prop}\label{PW}
We have
\begin{equation}\label{u2}
u_{\pm 1/2}(x;\xi = 1)   \mathop{\sim}\limits_{x \to \infty} \mp {x^{1/2} \over 2},
\end{equation}
 while for $0 < \xi < 1$ we have
 \begin{equation}\label{u3}
- 2^{-1/3}u_{\pm 1/2}(2^{-1/3}x;\xi)   \mathop{\sim}\limits_{x \to \infty} 
{\log (1 - \xi) \over 2 \pi} x^{1/2}
\cos^2 \Big ( {2 \over 3} x^{3/2} - {3 \over 4} d^2 \log x - c \Big ),
\end{equation}
where $d$ and $c$ are as in (\ref{N2}).
 \end{prop}
 
\noindent Proof. \quad The behaviour of $u_{1/2}(x;1)$ given in (\ref{u2}) follows by
substituting $\mu = 1/2$ in (\ref{B3}), then using (\ref{B1a}). With this established,
the   behaviour of $u_{-1/2}(x;1)$ follows after noting
from (\ref{b1}) that
\begin{equation}\label{kj}
u_{1/2}(x;\xi) - u_{-1/2}(x;\xi) = - 2^{1/3} q_0(-2^{1/3}x;\xi)
\end{equation}
and a further use of (\ref{B1a}). 

To derive (\ref{u3}) in the case $\mu = 1/2$, we simply make use of (\ref{B3}) with $\mu = 1/2$,
and then substitute (\ref{N2}).  The result for $\mu = -1/2$ now follows from (\ref{kj}) and
further use of  (\ref{N2}).  \hfill $\square$

\medskip
In distinction to the forms (\ref{b3}) and (\ref{b31}) for $\mu =1$ and 2, the dependence on $\xi$
for $\mu = \pm 1/2$ as exhibited in Proposition \ref{PW} appears already in the leading term, and
furthermore this term is, for $\xi \ne 1$, oscillatory. This latter feature can be understood from
the details of the average (\ref{A}). Thus for $\xi = 1$ the quantity being averaged is always positive
independent of $\mu$,
while for $0 \le \xi < 1$ and $\mu = \pm 1/2$ this quantity can take on complex values, but for
$\mu \in \mathbb Z^+$ and $\xi$ in this range it is again real. This suggests that in the variable 
$\xi^*$ as defined above (\ref{A}), with $0 < \xi^* \le 1$, the leading asymptotic form will be
analogous to that seen in (\ref{b3}) and (\ref{b31}), and that the $\xi$ dependence
will only be seen in a trans-series. As already noted in \cite[Eq.~(5.35)]{FW00}, this leading form is
expected to be
\begin{equation}\label{5.35}
u_{\mu}(x;\xi^*)  \mathop{\sim}\limits_{x \to \infty}  - \mu x^{1/2} - {\mu^2 \over 4 x} + {\mu(4\mu^2 + 1) \over 32 x^{5/2}} + \cdots
\end{equation}
According to (\ref{B3}) and  (\ref{kj})  this could be checked for $\mu = \pm 1/2$ upon knowledge of the $x \to - \infty$ form of $q_0(x,1\pm i (1-\xi^*))$, if this was to be available (see \cite{Cl08} for an easy to read review on what is available).

For general (but fixed) values of $\xi$, Pr\"ahofer and Spohn \cite{PS00} have shown how the Hasting-MacLeod solution $q_0(s;\xi)$ can be numerically computed to very high precision. According to formulas presented above, this means too that $u_{\mu}(x;\xi)$ for $\mu = \pm 1/2, 1$ and 2 can similarly be numerically computed. Moreover, with these values as initial conditions,
we can then compute $u_{\mu}(x;\xi)$ for general positive integer or half integer values by using the fact that the latter satisfy the alternate discrete Painlev\'e I equation \cite[Eq.~(5.12)]{FW00}
\begin{equation}\label{5.12}
{\mu \over u_{\mu +1}(x;\xi) - u_{\mu-1}(x;\xi)} +
{\mu + 1 \over u_{\mu +2}(x;\xi) - u_{\mu}(x;\xi)} = x - 
(u_{\mu +1}(x;\xi) - u_{\mu}(x;\xi))^2.
\end{equation}
We remark too that
this equation, in theory at least, allows the leading $\xi$-dependent terms in the $x \to \infty$ asymptotic expansion of $u_\mu(x;\xi)$ to be
determined for all $\mu = 0,1,2,\dots$. However in practice, as it appears from (\ref{ud}), (\ref{b3}) and (\ref{b31}) that the leading such
term is proportional to $\xi x^{- 1 - 3 \mu/2}  e^{-{4 \over 3} x^{3/2}}$, this effectively means expanding the trans-series of $u_0(x;\xi)$ to 
one higher order for each integer increase in $\mu$,  due to cancellations.

\subsection{Boundary spectrum singularity}\label{S33}
Closely related to (\ref{A}) is
 the PDF (\ref{Cg}) in the case $\beta = 2$ and $g(\lambda) = |\lambda - a|^\mu e^{-\lambda^2}$. For $\mu \in \mathbb Z^+$, this has the
interpretation as the eigenvalue PDF for the Gaussian unitary ensemble, conditioned so that the point $\lambda = a$ is an eigenvalue with
multiplicity $\mu$. It is well defined for general $\mu>-1$. 

The best known case of this PDF is when $a=0$. In \cite{NS93} the correlation functions for the scaled state defined by
replacing each $\lambda_l$ by $\pi \lambda_l/\sqrt{2N}$ and taking $N \to \infty$ were computed. In keeping with the interpretation for
$\mu \in \mathbb Z^+$, this state was subsequently referred to as the (bulk) spectrum singularity (see e.g.~\cite{Fo10}).
More recently, Its et.~al \cite{IKO08} considered the situation in which the spectrum singularity at $\lambda = a$ is located within the soft edge boundary
layer. (In fact in \cite{IKO08} the generalization of PDF with $e^{-\lambda^2}$ replaced by $e^{-V(x)}$, for any $V(x)$ inducing a soft edge, was
considered.)

In general the $k$-point correlation function for a unitary invariant ensemble ($\beta = 2$ case of (\ref{Cg})) is given by a determinant
\begin{equation}\label{kp}
\rho_{(k)}(x_1,\dots,x_k) = \det [ K_N(x_j,x_l) ]_{j,l=1,\dots,k},
\end{equation}
where the correlation kernel is given in terms of the orthogonal polynomials of one variable $\{p_n(x) \}_{n=0,1,\dots}$ with respect to the
weight function $g(x)$ according to
\begin{align}\label{5.1}
K_N(x,y) &  = (g(x) g(y) )^{1/2}  \sum_{j=0}^{N-1} {1 \over  {\mathcal N}_j}  p_j(x) p_{j}(y)
\nonumber \\
 & = (g(x) g(y) )^{1/2} {p_N(x) p_{N-1}(y) - p_N(y) p_{N-1}(x) \over {\mathcal N}_{N-1} (x-y) }
\end{align}
(see e.g.~\cite[Ch.~5]{Fo10}) with $\mathcal N_j := \int_{-\infty}^\infty g(x) (p_j(x))^2 \, dx$. The main result of \cite{IKO08} is the evaluation of the
scaled limit
$$
K^{\rm soft}_\mu (x,y;c) := \lim_{N \to \infty} {1 \over \sqrt{2} N^{1/6}}
K_N \Big ( \sqrt{2N} + {x \over \sqrt{2} N^{1/6}},   \sqrt{2N} + {y \over \sqrt{2} N^{1/6}} \Big ) \Big |_{a = \sqrt{2N} + c/(\sqrt{2} N^{1/6})}.
$$
The functional form 
\begin{align}\label{5.1a}
K^{\rm soft}_\mu (x,y;c) = {\psi_2(x;c) \psi_1(y;c) -  \psi_1(x;c) \psi_2(y;c) \over 2 \pi i (x-y) }
\end{align}
was obtained, where the $\psi_j$, given explicitly in \cite{IKO08}, can be expressed in terms of the Flaschka-Newell \cite{FN80} Lax pair components for the Painlev\'e II equation (\ref{1.1}) with
$\alpha = \mu + 1/2$. 

This characterization was shown to dramatically simplify in the cases $\mu = 0, 2$ to forms
involving only Airy functions,
\begin{align}\label{3.27}
K^{\rm soft}_0(x,y;c) & = {{\rm Ai}(x+c)  {\rm Ai}'(y+c)  -  {\rm Ai}'(x+c)  {\rm Ai}(y+c)  \over x - y}
\nonumber \\
K^{\rm soft}_2(x,y;c) & = {K^{\rm soft}_0(x,y;c) K^{\rm soft}_0(0,0;c) -  
K^{\rm soft}_0(x,0;c) K^{\rm soft}_0(y,0;c)  \over
 K^{\rm soft}_0(0,0;c) }.
\end{align}
These simplification were achieved by firstly demonstrating that the Flaschka-Newell
Lax pair can be written in terms of Airy functions for $\alpha = 1/2$.

The functional form for $K_0^{\rm soft}(x,y;c)$ is the fundamental Airy kernel (in the variables
$x+c$, $y+c)$) which underlies (\ref{Cg2}). The corresponding $k$-point correlation function
$\rho_{(k)}^{(\mu = 0,c)}$ is given by substituting this for $K_N$ in (\ref{kp}). On the other hand,
we see from the definitions that for $m=0,1,\dots$
\begin{equation}\label{pp}
\rho_{(k)}^{(2m+2,c)}(x_1,\dots,x_k) =
\lim_{x_{k+1} \to 0}
{ \rho_{(k+1)}^{(2m+2,c)}(x_1,\dots,x_k,x_{k+1};c) \over
\rho_{(1)}^{(2m,c)}(x_{k+1}) }
\end{equation}
(the case $m=0$ of this formula was used in \cite{FO96} in relation to the bulk spectrum singularity ensemble). This allows us to express $K_{2m+2}^{\rm soft}$ in terms of $K_{2m}^{\rm soft}$,
thus generalizing the second of the formulas in (\ref{3.27}).

\begin{prop}\label{C5}
We have
\begin{equation}\label{3.28}
K^{\rm soft}_{2m+2}(x,y;c)  = 
\lim_{z \to 0} {K_{2m}^{\rm soft}(x,y;c)  K_{2m}^{\rm soft}(z,z;c)  -  
K_{2m}^{\rm soft}(x,z;c) K_{2m}^{\rm soft}(z,y;c) \over K_{2m}^{\rm soft}(z,z;c) }.
\end{equation}
\end{prop}

\noindent
Proof. \quad We have, for small $x_{k+1}$,
\begin{eqnarray*}
\lefteqn{
 \rho_{(k+1)}^{(2m+2,c)}(x_1,\dots,x_k,x_{k+1};c) } \\
&& := \det \begin{bmatrix} [K_{2m}^{\rm soft}(x_j,x_l;c)]_{j,l=1,\dots,k} &
 [K_{2m}(x_j,0;c)]_{j=1,\dots,k} \\
 [K_{2m}^{\rm soft}(0,x_k;c)]_{l=1,\dots,k} & K_{2m}(x_{k+1},x_{k+1};c)
 \end{bmatrix} \nonumber \\
 && \sim
 \det \begin{bmatrix} [K_{2m+2}^{\rm soft}(x_j,x_l;c)]_{j,l=1,\dots,k} &
 [0]_{j=1,\dots,k} \\
 [0]_{l=1,\dots,k} & K_{2m}^{\rm soft}(x_{k+1},x_{k+1};c)
 \end{bmatrix},
 \end{eqnarray*}
 where the second equality follows by applying elementary row and column operations to the
 first determinant., and taking $x_{k+1} \to 0$ in the first block. Expanding by the last row, and noting that
 $$
 \rho_{(1)}^{(2m,c)}(x_{k+1}) =  K_{2m}^{\rm soft}(x_{k+1},x_{k+1};c)
 $$
 gives (\ref{3.28}). \hfill $\square$
 
 \medskip
 It is a basic but fundamental fact that the gap probability in a statistical mechanical system
 can be expressed as a sum over the correlation functions (see e.g.~\cite[Eq.~(9.4)]{Fo10}).
 For the boundary spectrum singularity
 ensemble, the generating function for the probability of exactly $k$ ($k=0,1,2,\dots$)
 eigenvalues in the
 interval $(0,\infty)$ is precisely the ratio
 $$
 {E_2^{\rm soft}(c;\mu;\xi) \over  E_2^{\rm soft}(c;\mu;\xi=0) },
 $$
 where $E_2^{\rm soft}(c;\mu;\xi)$ is as specified in Section \ref{S33}. Using the expansion of
 this generating function in terms of the correlations one has
 \begin{equation}\label{s33}
 {E_2^{\rm soft}(s;\mu;\xi) \over  E_2^{\rm soft}(s;\mu;\xi=0) } \mathop{\sim}\limits_{c \to \infty}
 1 - \xi \int_0^\infty \rho_{(1)}^{(\mu,c)}(x) \, dx + \cdots
\end{equation}
Recalling (\ref{up}), it follows that
\begin{equation}\label{s34}
u_\mu(c;\xi) - u_\mu(c;\xi=0)   \mathop{\sim}\limits_{c \to \infty}
 - \xi {d \over dc}  \int_0^\infty \rho_{(1)}^{(\mu,c)}(x) \, dx.
 \end{equation}
 
 In the case $\mu = 0$, (\ref{s34}) is equivalent to  (\ref{ud}). Moreover, we can use (\ref{s34})
 to reclaim the $\xi$ dependent term in (\ref{b31}). Thus a straightforward calculation (peformed
 using computer algebra), expanding $K^{\rm soft}_2(x,y;c) $ as specified by (\ref{3.27}) for
 large $c$ shows that
 $$
 K^{\rm soft}_2(x,y;c)  \mathop{\sim}\limits_{c \to \infty} e^{-(4/3) c^{3/2}}
 e^{-(x+y) \sqrt{c}} {xy \over 2^7 \pi c^3}.
 $$
 Recalling that $ \rho_{(1)}^{(\mu=2,c)}(x)  = K^{\rm soft}_2(x,y;c) $ we substitute this in
 (\ref{s34}) to conclude
 \begin{equation}\label{s35}
u_\mu(c;\xi) - u_\mu(c;\xi=0)   \mathop{\sim}\limits_{c \to \infty}  \xi {e^{-(4/3) c^{3/2}} \over 256 \pi c^{4}},
\end{equation}
which is in precise agreement with (\ref{b31}). 

We can can furthermore demonstrate that all terms
proportional to $\xi$ as given by (\ref{s34}) agree with those obtained from (\ref{b2}).
For this we must extend the first asymptotic equality in (\ref{ud}) to read
\begin{equation}\label{u0d}
u_0(c;\xi) \mathop{\sim}\limits_{c \to \infty} \xi K^{\rm soft}_0(0,0;c) - \xi^2 \int_0^\infty (K^{\rm soft}_0(x,0;c))^2 \, dx + O(\xi^3).
\end{equation}
This is a consequence of (\ref{up}) and the fact that $E_2^{\rm soft}(s;0;\xi) = \det (1 - \xi K_{(s,\infty})$, where
$K_{(s,\infty)}$ is the integral operator on $(s,\infty)$ with kernel $ K^{\rm soft}_0(x,y;0) $, and the expansion of the latter in
terms of determinants (or equivalently the underlying correlation function; essentially we are extending the case $\mu = 0$
of (\ref{s33}) to second order --- as an aside we remark that this formula provides an alternative method to derive
the coefficients in the expansion (\ref{dd0})). Substituting (\ref{u0d}) in (\ref{b2}) reclaims,  after minor manipulation,
(\ref{s34}).

\subsection*{Acknowledgements}
 This work was supported by the Australian Research Council.
 

\providecommand{\bysame}{\leavevmode\hbox to3em{\hrulefill}\thinspace}
\providecommand{\MR}{\relax\ifhmode\unskip\space\fi MR }
\providecommand{\MRhref}[2]{%
  \href{http://www.ams.org/mathscinet-getitem?mr=#1}{#2}
}
\providecommand{\href}[2]{#2}

\end{document}